
%
%
%
%
\input harvmac
\noblackbox
\def\Title#1#2{\rightline{#1}\ifx\answ\bigans\nopagenumbers\pageno0\vskip1in
\else\pageno1\vskip.8in\fi \centerline{\titlefont #2}\vskip .5in}

%
%
\def\tpm{T_{+-}}
\def\tpp{T_{++}}
\def\tmm{T_{--}}
\def\emtf{e^{-2\phi}}
\def\hf{{1\over2}}
\def\fdp{\phi^{\prime\prime}}
\def\rdp{\rho^{\prime\prime}}
\def\etr{e^{2\rho}}
\def\Nfe{{N\over48}}
\def\ppl{\partial_+}
\def\pmi{\partial_-}
\def\etf{e^{2\phi}}
\def\LDV{linear dilaton vacuum}
\def\emts{e^{-2\lambda\sigma}}
\def\Ntf{{N\over24}}
\def\calO{{\cal O}}
\def\emfs{e^{-4\lambda\sigma}}
\def\phic{\phi_{\rm cr}}
\def\emf{e^{-\phi}}

\def\xp{x^+}
\def\xm{x^-}
\def\Ntw{{N\over12}}
\def\that{{\hat t}}
\def\etfo{e^{2 \phi(0)}}
\font\ticp=cmcsc10
\def\hf{{1\over2}}

\def\ajou#1&#2(#3){\ \sl#1\bf#2\rm(19#3)}

\def\frac#1#2{{#1 \over #2}}

\lref\BDDL{T. Banks, A. Dabholkar, M.R. Douglas, and M O'Loughlin, ``Are
horned particles the climax of Hawking evaporation?'' Rutgers preprint
RU-91-54.}
\lref\caru{ C.~Callan, ``Disappearing dyons,''\ajou Phys. Rev. &D25 (82)
2141; ``Dyon - fermion dynamics,''\ajou  Phys. Rev.
&D26 (82) 2058; ``Monopole catalysis of baryon decay,''\ajou Nucl. Phys.
&B212 (83) 391\semi
V. Rubakov, ``Superheavy magnetic monopoles and proton decay,''\ajou
Pis'ma Zh. Eksp. Teor. Fiz. &33 (81) 658 (\ajou JETP
 Lett. &33 (81) 644); ``Adler-Bell-Jackiw anomaly
and fermion number breaking in the presence of a magnetic monopole,''\ajou
Nucl. Phys. &B203 (82) 311.}
\lref\dxbh{S.B. Giddings and A. Strominger, `` Dynamics of
extremal black holes'', UCSB preprint UCSB-TH-92-01, hepth@xxx/9202004,
to appear in {\sl Phys. Rev.} {\bf D}.}
\lref\RST{J.G. Russo, L. Susskind, and L. Thorlacius, ``Black hole
evaporation in 1+1 dimensions,'' Stanford preprint SU-ITP-92-4.}
\lref\HoWi{C.F.E. Holzhey and F. Wilczek, ``Black holes as elementary
particles,'' IAS preprint IASSNS-HEP-91/71.}
\lref\WittSton{E.~Witten, ``On black holes and string theory,'' IAS
preprint (Lecture
notes from Strings and Symmetries '91, Stony Brook).}
\lref\alf{M. Alford and A. Strominger, to appear.}
\lref\evan{C. Callan, S.B. Giddings, J. Harvey and A. Strominger,
``Evanescent Black Holes''
{\sl Phys. Rev.} {\bf D 45} (1992) 1005.}
\lref\Hawk{S. W. Hawking, ``Particle creation by black holes,''
\ajou Comm. Math. Phys. &43 (75) 199.}
\Title{\vbox{\baselineskip12pt
\hbox{UCSB-TH-92-08}\hbox{EFI-92-16}
\hbox{hepth@xxx/9203042}}}
{Quantum Black Holes}
\baselineskip=12pt
\bigskip
\centerline{\ticp Bjorn Birnir,$^*$ Steven B.
Giddings,$^{\dagger}$}
\medskip
\medskip
\centerline{\ticp Jeffrey A. Harvey,$^\#$ and Andrew
Strominger$^{\dagger}$}
\medskip
\bigskip
\centerline{\bf Abstract}

Static solutions of large-$N$ quantum dilaton gravity in $1+1$
dimensions are analyzed and found to exhibit some unusual
behavior. As expected from previous work, infinite-mass solutions are found
describing a black hole in equilibrium with a bath of Hawking radiation.
Surprisingly, the
finite mass solutions are found to approach zero coupling
both at the horizon and spatial infinity,
with a ``bounce'' off of strong coupling in between.
Several new zero mass solutions --  candidate quantum vacua -- are
also described.
\bigskip
\bigskip
\bigskip
\bigskip
\bigskip
\medskip
{\sl $^*$Department of Mathematics, University of California,} {\sl Santa
Barbara, }

{\it CA 93106 Internet: birnir@henri.ucsb.edu}

{\sl$^\dagger$Department of
Physics, University of California,} {\sl Santa Barbara, CA 93106}

{\it
Internet: giddings@denali.physics.ucsb.edu, andy@denali.physics.ucsb.edu}


{$\#$\sl Enrico Fermi Institute, University of Chicago, 5640 Ellis Avenue,

Chicago, IL 60637 }
{\it Internet: harvey@yukawa.uchicago.edu}

\Date{3/92}

\newsec{ Introduction}  

In his seminal work \Hawk\ Hawking argued that the laws of quantum
mechanics, when applied to black holes, predict their own demise: a pure
state which collapses into a black hole evaporates into a mixed final
state.
In the intervening fifteen years,
progress in verifying or refuting his claims has been stymied by several
formidable obstacles. One of these is that regions of Planck scale curvature
and strongly coupled quantum gravity probably arise in four-dimensional
gravitational collapse.  While string theory provides a model for
describing
weakly coupled quantum gravity, a description of
strongly coupled quantum gravity is well
beyond our reach. Another obstacle is the problem of analyzing the
backreaction of
Hawking radiation on the gravitational field.  There are
indications are that this
is qualitatively important in resolving the puzzle of information loss,
yet practical methods for describing it have not been forthcoming.

Recently, a strategy for sidestepping the first obstacle and overcoming
the second was proposed \evan. A great simplification
occurs by considering the problem of black hole formation/evaporation
in a renormalizable theory of ``dilaton'' gravity in ${1}{+}{1}$ dimensions
coupled to conformal matter. This ``toy'' problem contains most of
the important
conceptual issues present in the four-dimensional case, yet is computationally
much more tractable. The region of strongly coupled quantum gravity can be
analyzed within the framework of a ${1}/{N}$ expansion where ${N}$ is the
central charge of the conformal matter. The problem of black hole
formation/evaporation, including the gravitational backreaction, is
thus formally reduced to a system of second order partial differential
equations. In addition to serving as a two-dimensional model for
four-dimensional gravitational collapse, this two-dimensional
theory is also directly relevant to four-dimensional
physics as the effective
theory describing the absorption/re-emission of incident  particles by
certain extremal dilatonic black holes in four dimensions \refs{\evan
\BDDL-\dxbh}.

The analysis of \evan\ expanded the theory around the
``linear dilaton vacuum'' configuration. This is a static solution
of the large ${N}$ equations of motion for which the dilaton varies linearly
across space. Since the dilaton governs the strength of quantum loops,
quantum fluctuations are large in half of space (referred to
as the ``Liouville region'') and small in the other half (the
``dilaton region''). There is a sharp line dividing these
two regions along which the dilaton take the critical value ${\phi} =
{\phic}$. A black hole is potentially formed by sending matter in from the
dilaton region to the Liouville region. While the equations describing
this process were not solved, it was conjectured in \evan\
that the collapsing
matter loses all its energy via Hawking radiation before the black hole
has a chance to form.

This conjecture was shown in \refs{\BDDL,\RST} to be false. In fact something
rather different occurs; when the collapsing matter tries to cross
${\phic}$ from the dilaton to the Liouville region, a singularity
appears \refs{\BDDL,\RST}. This singularity is quite different in
nature from the black hole singularities of the classical theory: it
occurs at the finite value ${\phi} = {\phic}$ (as
opposed to the strong coupling value ${\phi} = {\infty}$,) and the metric
also remains finite at the singularity.
Its physical
significance is somewhat mysterious.

In an attempt to better understand the nature of the
critical line ${\phi} = {\phic}$ separating the two phases, and the physical
implications of the singularities, in this paper we investigate
numerically and analytically the static
solutions of the large-$N$ equations. We find that there is
a rich variety of solutions with some
rather unexpected behavior: there is a tendency to ``bounce'' off of the
critical line. We also find vacuum solutions with greater symmetry than the
\LDV\ which lie wholly within the Liouville region, as well as a zero mass
black hole with a singular horizon which lies entirely within the dilaton
region.

After a brief summary of the relevant formula in section 2, some static
solutions are discussed in section 3. We also argue there that the endpoint
of evaporation of black holes formed in the \LDV\ is a truncated \LDV\
terminated just before a singularity at $\phic$. In section 4 we argue
that the singularities at $\phic$ signal a breakdown of the ${1 \over N}$
expansion, and are conceivably resolved in the exact quantum theory.
The related possibility of defining the large-$N$ theory by boundary
conditions at $\phic$ is also discussed.

\newsec{Summary of Relevant Previous Results}

Classical dilaton gravity coupled to $N$ conformal
matter fields is described by the action
\eqn\one
{S= { 1 \over 2\pi}\int d^2 \sigma\sqrt{-g}\Bigl[e^{-2\phi}(R+4(\nabla\phi)^2
+4\lambda^2)
-\half\sum\limits^N_{i=1}(\nabla f_i)^2\Bigr],}
where $g$, $\phi$ and $f_i$ are the metric, dilaton, and matter fields,
respectively, and $\lambda^2$ is a cosmological constant.  The matter
fields can be explicitly integrated out to provide an effective action for
the metric and dilaton.  In the conformal gauge this action is
%
\eqn\ntythree{\eqalign{S_N &= {1 \over \pi}\int\ d^2\sigma\ \Bigl[e^{-2\phi}
\Bigl( \partial_+(2\phi-\rho)\partial_-(2\phi-\rho)
-\lambda^2 e^{2\rho}\Bigr)\cr
             & + ({N \over 12}-e^{-2\phi} )
\partial_+\rho\partial_-\rho\Bigr]~,\cr}}
where we have chosen the conformal gauge
\eqn\two
{\eqalign{g_{+-} &=-\half e^{2\rho},\cr
          g_{--} &= g_{++} = 0.\cr}}
The term proportional to $N$ is the Liouville term induced by the matter
fields. The equations of motion for $\rho$ and  $\phi$ are
\eqn\rheq{\eqalign{T_{+-}  &= e^{-2\phi}(2\partial_+\partial_-\phi - 4
\partial_+\phi\partial_-\phi - \lambda^2e^{2\rho})\cr
                   &- {N \over 12} \partial_+\partial_-\rho\ =0,\cr}}
\eqn\four{-4\partial_+\partial_-\phi + 4\partial_+\phi\partial_-\phi +
2\partial_+\partial_-\rho + \lambda^2 e^{2\rho}=0.}
%
%
%
Because the action  \ntythree\ is gauge fixed, these equations
of motion should be supplemented by the constraint
\eqn\ntyfour
{\eqalign{T_{++}& = e^{-2\phi} (4\partial_+\phi\partial_+\rho -
2\partial^2_+\phi) \cr
                  & - {N \over 12}\left(\partial_+\rho\partial_+\rho -
\partial^2_+\rho + t_+(\sigma^+)\right)=0,\cr}}
as well as a similar equation for $T_{--}$.\foot{However we note that
the Bianchi identity implies that any solution of \rheq\ - \four\
is also a solution of the constraints for some choice of $t_\pm$.} $t_+$ is an
integration function which must be fixed by boundary conditions.
Solving these large-$N$
equations includes the effects of Hawking radiation as well as the
gravitational backreaction.

     An important solution of these equations is known as the linear
dilaton vacuum:
\eqn\ntyfive{\eqalign
{\rho&=0,\cr
  \phi&=-\frac{\lambda}{2} (\sigma^+-\sigma^-).\cr}}
This vacuum is divided into two regions by the critical line
\eqn\phcr{\phi=-\half\ln{N \over 12}\equiv \phi_{\rm cr},}
across which, as easily seen from \ntythree, an eigenvalue of the
kinetic operator changes sign. Quantum effects are
small in the region $\phi<\phi_{\rm cr}$, which is referred to as the
dilaton region. The gravitational dynamics are governed by the
Liouville action in the region $\phi>\phi_{\rm cr}$, which is referred to
as the Liouville region.

\newsec{Static solutions}


In this section we shall describe some static solutions of
equations \rheq\ -- \ntyfour.
In the static limit these equations become
\eqn\static{\eqalign{0&=\tpm= \emtf \left( - \hf \fdp + \phi^{\prime 2} -
\lambda^2 \etr\right) + \Nfe \rdp, \cr
0&=\tpp=\tmm=\emtf\left(\phi' \rho' -\hf \fdp\right) - \Nfe
\left(\rho^{\prime 2} -\rdp +t \right), \cr
0&={\delta S\over \delta \phi} \propto \fdp - \phi^{\prime 2} -\hf \rdp +
\lambda^2 e^{2\rho}, }}
where $t$ is a constant and prime denotes $d/d\sigma$, with $\sigma=
\half (\sigma^+-\sigma^-)$.  These equations are of
course redundant; the $\tpm$  and dilaton equations imply the vanishing of
$(\tpp - \tpm)'$.  The $\tpm$ and dilaton equations can also be rewritten in
the form
\eqn\nstatic{\eqalign{&\left( 1-{N \etf \over 24} \right) \rdp =\fdp , \cr
&\hf \left(1-{N \etf \over12}\right) \fdp = \left(1-{N \etf \over 24}
\right) \left( \phi^{\prime 2} - \lambda^2 \etr\right),}}
which will be convenient for later use.

\subsec{Quantum Kinks}
Solutions to equations \static\ may be specified by fixing $\rho$, $\phi$, and
their derivatives at infinity. The mass of a solution asymptotic to
\ntyfive\ is
\eqn\mss{M=2e^{2\lambda\sigma}(\lambda\delta\rho+\delta\phi ')}
evaluated at infinity
where $\delta\rho,~\delta\phi$ are the deviations from \ntyfive.
We are in particular interested in
solutions asymptotic to the linear dilaton vacuum, with finite mass and
with no incoming or outgoing energy flux at infinity.  These are candidates
for the final state of black hole evaporation.
The asymptotic behavior of such solutions can be determined by linearizing
\static\ around the \LDV\ in the gauge \ntyfive.
The resulting equations can be put in the form
\eqn\linear{\eqalign{ 2 \lambda\delta\phi'& + 2 \lambda^2\delta\rho -
\lambda
\delta\rho' =
\Nfe\emts t,\cr \delta \fdp& = \delta\rdp \left(1-\Ntf \emts\right). }}
Solving \linear\ for the asymptotic perturbations of the \LDV\ yields an
infinite mass solution for $t\neq0$.  This makes physical sense:  $t\neq0$
corresponds to a constant incoming and outgoing energy flux, and the
solution has divergent ADM mass.  Such solutions, while perhaps interesting
for other reasons, are therefore not candidates for the final state of
black hole evaporation.

With $t=0$, the asymptotic solutions are of the form
\eqn\asymp{\eqalign{\delta\rho& = -{ M \over 2 \lambda}\emts +
\calO\left(\emfs\right)\cr
\delta\phi& = -{M \over 2 \lambda} \emts + \calO \left(\emfs\right) }}
and have finite mass $M$.  To find the geometry of these solutions one can
integrate the equations \static\ in from infinity.
$t=0$ defines an invariant surface of fixed $\tpp -\tpm$.  Different orbits
are chosen by initial data on this surface as illustrated in
figures 1a through 1d.  These are
solutions with distinct masses that asymptotically approach the linear
dilaton vacuum.

These solutions may be qualitatively understood as follows.  Asymptotically
they approach the \LDV\ plus the perturbations in \asymp.  Integrating in
from infinity, for positive
mass one finds from \nstatic\ that $\fdp<0$ and so they begin to turn over.
$\fdp$ becomes very large and negative in the
critical region $\phi \simeq \phic = -\hf \ln(N/12)$.  In this
vicinity an approximate solution can be found by setting
$\rho=$constant$=0$ and $\phi = \phic + \varphi$, where $|\varphi|\ll
|\phic|$.  (The consistency of this can then be checked using the full
equations.)  The resulting equation is
\eqn\perteq{ - \varphi^{\prime\prime} \varphi \simeq \hf \left(
\varphi^{\prime 2} -\lambda^2 \right),}
with first integral
\eqn\fiint{\varphi^{\prime 2} - {A\over \varphi} \simeq \lambda^2,}
where $A$ is an integration constant.  For positive mass, $A>0$ and the
equation is that for motion of a particle in a repulsive potential centered
at $\varphi=0$ (note that $\varphi<0$).  Hence $\phi $
bounces near $\phi=\phic$ ($\phi '$ flips from roughly
$-\lambda$ to $+\lambda$) and then begins to decrease towards $\sigma=-\infty$.
The sharp turnover in $\phi$ also forces $\rho '$ to jump, by
\nstatic.  $\rho$ is then found to be asymptotically linear and approaches
minus infinity.

The qualitative behavior as $\sigma\rightarrow -\infty$ can also be
understood, using the fact that $\etr$ becomes negligible as
$\rho\rightarrow - \infty$.  Without this term the equations can be
integrated to find
\eqn\integ{\eqalign{\emtf& = -{N\over12}\rho + a\sigma +b,\cr
{\emf}&\sqrt{ \emtf -\Ntw} - {\Ntw} \ln
\left[\sqrt{ \emtf -\Ntw} +  \emf \right] = -a\sigma +c,}}
where $a,b$, and $c$ are constants.  These have asymptotic solutions as
$\sigma\rightarrow -\infty$
\eqn\asympsoln{\eqalign{\emtf &\sim -a \sigma + \Ntf \ln ( -a\sigma
) + \cdots ,\cr
ds^2 &\sim { e^{48a\sigma/N} \over -a\sigma }(1 + \cdots) (-d\tau^2 +
d\sigma^2)\ .}}
One sees that $\sigma=-\infty$ is in fact an event horizon at finite
distance, and its vicinity is more easily investigated by introducing the
new coordinates
\eqn\newcoor{\xp=e^{24a\sigma^+/N}\ ,\ \xm = -e^{-24a\sigma^-/N}\ .}
{}From \asympsoln\ one finds infinite curvature at  $\xp=0$ and
$\xm=0$; the horizon is singular.  The singularity occurs at zero
coupling.

As the mass goes to zero, the solution gets
closer and closer to $\phic$ before bouncing back to weak coupling.
Furthermore, it is very close to the \LDV\ outside a region whose boundary
gets closer to $\phic$. A configuration can be defined in
the zero mass limit which agrees with the \LDV\ up to $\phic$, but then
bounces back to weak coupling ($\phi=-\infty$) rather than continuing
on to strong coupling ($\phi=+\infty$). The existence of distinct
``solutions'' which agree up to $\phic$ but then disagree afterwards is due
to the fact that the equations themselves degenerate at $\phic$. To
resolve this ambiguity one must go beyond the large-$N$ limit, as
will be discussed
in the last section.

Is this zero mass bounce
solution a plausible endpoint for black hole evaporation?
We think not. As described in \refs{\RST,\BDDL}, the black holes formed
by $f$-wave collapse have a dilaton which increases monotonically
up to a singularity at $\phic$. The static solutions described here are
non-singular at $\phic$: rather they bounce off of $\phic$ and reach a
singularity at $\phi=-\infty$. It is hard to see how the black holes
formed in a collapse process
could smoothly evolve into such a configuration.

It is tempting to try to instead interpret the zero mass bounce solution
as the true quantum vacuum of the theory. The singularities described in
\refs{\RST,\BDDL} might then be viewed as punishment for expanding
around the wrong vacuum. Making sense of this idea would require
finding
some sensible choice of boundary conditions at the horizon, as well
as for propagating through the kink at $\phic$; we have done neither.

\subsec{Quantum Black Holes}
The conditions for solutions with regular horizons are most easily
investigated by introducing a new spatial coordinate $s=-\xp\xm$, so that the
horizon is at $s=0$.  In terms of this coordinate the static equations
become
\eqn\seqns{\eqalign{&-2\phi' -2s\fdp + 4s \phi^{\prime 2} -\lambda^2
 \etr = -\Ntw \etf
\left(s \rdp + \rho'\right)\quad (+-)\cr
&4\phi'\rho'-2\fdp = \Ntw\etf\left(\rho^{\prime 2} -\rdp + {\that \over
s^2}\right)\quad (++,--)\cr & 4\phi' +4s\fdp -4s \phi^{\prime 2} -2 \rho' -2s
\rdp +\lambda^2\etr =0\quad ({\rm Dilaton}). }}
The conditions for the solution to be regular at the horizon then
follows from finiteness of $\fdp$ and $\rdp$ or, equivalently, the
vanishing of $s\fdp$ and $s\rdp$ at $s=0$. One finds
\eqn\rhor{\eqalign{\rho'(0)& = -{\lambda^2 \over 2} {e^{2\rho(0)}\over 1-\Ntw
\etfo},\cr
\phi'(0)& = -{\lambda^2 \over 2}
e^{2\rho(0)} {1-\Ntf \etfo \over 1- \Ntw \etfo},\cr
\that&=0\ .
}}
When translated into $\sigma$ coordinates the last condition implies that
$t_\pm=-\lambda^2/4\neq0$. In the $s$ coordinates $t=0$ generically implies
non-zero Hawking flux.
This means that the solutions with regular horizons only remain static
when supported by an incoming flux that matches the outgoing Hawking flux,
as expected.
This non-zero radiation density extending out to infinity implies that
these solutions
have infinite mass.
Numerical plots of these solutions can be found in figures 2a,b.
They can be continued across
the horizon, and a singularity appears at $\phi=\phic$.  The solutions
therefore have causal structures identical to the classical black hole.
Although they are not candidates for the final state, they approximate a
slowly evaporating black hole before it reaches zero mass.
\subsec{The Endpoint of Black Hole Evaporation}
To summarize the results up to this point, we have found static solutions
of two types:

\item{1.}{\it Solutions with $0<M<\infty$.  These have horizons with weak
coupling singularities.}

\item{2.}{\it Black holes with regular horizons.  These are solutions with
constant incoming radiation to balance the outgoing Hawking radiation, and
thus have infinite mass.}

Neither of these two types of solutions are good candidates for the final
state.  The only remaining candidate is at $M=0$:  the linear dilaton
vacuum.  However, because of the singularity at $\phi=\phic$, one should
terminate the \LDV\ just to the right of $\phic$.  One therefore presumes
that in the far future the solution formed from infalling matter
settles down to the \LDV\ for
$\phi$ in the range less than $\phic -\epsilon$, for some small $\epsilon$
which approaches zero in the infinite future. Thus the best candidate for
the  endpoint
of black hole evaporation may be pictured as the \LDV\ truncated
just before the end of the universe at $\phic$.

\subsec{The Liouville Region.}
In the preceding, static solutions which approached weak coupling at spatial
infinity were discussed. It is also of interest to consider solutions which
remain wholly within the Liouville region. Since the main problems are
associated with the crossover between the two regions, perturbation theory
around such solutions may well be better defined. Furthermore, as argued
in \refs{\BDDL,\dxbh}, such solutions are physically
relevant to the description of four-dimensional extremal black holes
for which the asymptotic value $\phi_0$ of the dilaton obeys
$\phi_0 >\phic$.

The simplest such solution is the vacuum configuration
\eqn\lvac{\eqalign{
{e^{-2\phi}}&={0},\cr
{\rho}&={0}.\cr}}
Perturbation theory about this vacuum can be defined in terms
of the variable\foot{We require ${\psi}$, but not ${\phi}$,
to be real.}
\eqn\psidef{
{\psi}\,\,{\equiv}\,\,{e^{-\phi},}}
in terms of which the large ${N}$ gravitational action becomes:
\eqn\psiact{
\eqalign{
{S}&= {{1}\over{\pi}}\,\,{\int}\,\,{d^2}{\sigma} ({4}{\partial}_+{\psi}
{\partial}_-{\psi} + {4}{\psi}{\partial}_+{\psi}{\partial}_-{\rho}\cr
&\qquad - {\lambda^2}{\psi^2}{e^{2\rho}}
 + {{N}\over{12}}\,\,{\partial}_+{\rho}{\partial}_-{\rho}).\cr}}
Excitations about this vacuum are characterized by the conserved
energy:
\eqn\lien{
\eqalign{
{E}&= {-\hf}\,\,{\int}\,\,{d}{\sigma}\,\,
\biggl[({2}{\partial}_+{\psi} + {\psi}{\partial}
_+{\rho})^2 + ({2}{\partial}_-{\psi} + {\psi}{\partial}_-{\rho})^2\cr
&\qquad + {2}{\lambda^2}{\psi^2} {e^{2\rho}} +
\biggl({{N}\over{12}} - {\psi^2}\biggr)\,\,
\biggl(({\partial}_+{\rho})^2 + ({\partial}_-{\rho})^2\biggr)
+\Ntw(t_++t_-)\biggr].\cr}}
$E$ may also be expressed as a surface integral:
\eqn\esrf{E=-\hf\biggl[\Ntw (\ppl -\pmi )\rho+2\psi(\ppl -\pmi
)\psi\biggr]^{+\infty}_{-\infty},}
from which it is evident that $E=0$ for configurations asymptotic to \lvac.
A non-zero conserved quantity
\eqn\lien{
\eqalign{
{\hat E}&= \,\,\hf{\int}\,\,{d}{\sigma}\,\,
\biggl[({2}{\partial}_+{\psi} + {\psi}{\partial}
_+{\rho})^2 + ({2}{\partial}_-{\psi} + {\psi}{\partial}_-{\rho})^2\cr
&\qquad + {2}{\lambda^2}{\psi^2} {e^{2\rho}} +
\biggl({{N}\over{12}} - {\psi^2}\biggr)\,\,
\biggl(({\partial}_+{\rho})^2 + ({\partial}_-{\rho})^2\biggr)
\biggr].\cr}}
can also be defined by virtue of the fact that $t_\pm$ are functions only
of $x^\pm$. Note that for
${\psi^2}<{{N}\over{12}}$, which defines the Liouville region, $\hat E$
is positive semi-definite which constrains
possible choices of of $t_\pm$.

There is also a static solution corresponding to anti- deSitter space
with a constant dilaton.
It is easy to see directly from \rheq\ and \four\ that the
field configuration
\eqn\desit{\eqalign{\emtf &= \Ntf,\cr
R&=8e^{-2\rho}\partial_+\partial_-\rho=-4\lambda^2, \cr}}
is a solution. In static coordinates $\rho$ is given by
\eqn\rdes{\rho=-\ln ( \sqrt 2 \lambda\sigma).}

\newsec{Discussion}

In \refs{\RST,\BDDL} it was argued in the quantum theory that a small
perturbation of the linear
dilaton vacuum produces a black hole. The black hole then evaporates
leaving in its place a configuration close to the linear dilaton vacuum
until very near ${\phi} = {\phic}$, at which point there is a singularity.
Thus the linear dilaton vacuum is unstable under small perturbations, and
is in this sense not the true vacuum of the theory. A candidate for the
true vacuum is the zero-mass configuration with a singularity at ${\phi}
 = {\phic}$. One should endeavor to understand this configuration.

Can we really reliably conclude that there is a singularity at ${\phi}
= {\phic}$? The answer to this is no, because the ${1}/{N}$ expansion
breaks down before the singularity is reached, and the equations used
to find the singularity are not a good approximation to the quantum theory
described by \one. In terms of Feynman diagrams for perturbation theory
about the linear dilaton vacuum, the large-${N}$ action \ntythree\
describes graviton-dilaton tree diagrams plus one loop matter.
Graviton-dilaton loops are suppressed as long as the propagator is of order
${1}/{N}$. Equivalently, the determinant of the matrix ${K}$ governing
small fluctuations of, ${\rho}$, ${\phi}$ should be of order ${N}^2$. That
determinant is given by
\eqn\kdet{
{\rm det}\,\,{K} = {e^{-2\phi}}\,\,\biggl({e^{-2\phi}} -
{{N}\over{12}}\biggr).}
This is indeed of order ${N^2}$ in both the Liouville and dilaton
regions for ${e^{-2\phi}}$ of order ${N}$, {\it except} when ${e^{-2\phi}}$
is near ${{N}\over{12}}$, which demarcates the two regions of the
linear dilaton vacuum. Since
${\phi} =- {\lambda}{\sigma}$, the ${1}/{N}$ expansion breaks down in
a region of width of order ${\lambda^{-1}}$ containing the critical line or,
equivalently wherever $\phi$ differs from $\phic$ by an amount less than one.
Large ${N}$ fails to suppress quantum fluctuations of the dilaton and
metric within this region.

In a sense this brings us back to where we started from; the interesting
physics occurs in a region outside the reach of perturbation theory. Large
${N}$ has failed to fully tame strongly coupled quantum gravity. However,
some small progress has nevertheless been made in the following sense. The
breakdown of large-${N}$ perturbation theory is limited to a small
region of width ${\lambda^{-1}}$ (in contrast to ordinary perturbation
theory, which is bad in half of spacetime). If
we restrict our attention to processes on scales large compared to ${\lambda
^{-1}}$,\foot{This is in any case necessary if the two-dimensional
theory is viewed as an effective theory for four-dimensional black holes.}
this region is effectively a one dimensional line.

Progress might then be made if one {\it assumes} that the exact quantum
theory has a well-defined evolution. (Of course we have no evidence
in favor of this; it is important and difficult to find out
if this is indeed the case.) This exact quantum theory will then imply
some effective boundary conditions along the critical line in the
effective theory at scales larger than ${\lambda^{-1}}$. Constraints
on these boundary conditions can be derived from consistency of the low energy
theory.\foot{This is reminiscent of the problem of fermion-monopole
scattering analyzed by Callan and Rubakov \caru.} In the following
we consider two possible types of boundary conditions.

\subsec{ Bouncing off the singularity.}

In this picture the universe ends at ${\phi} = {\phic}$ (or just before
there,
so that the large-$N$ equations are valid), and boundary
conditions must be imposed there. In the vacuum, this line is timelike
(if it is drawn just before $\phic$), so
it might appear sensible to apply Dirichlet or Neumann boundary conditions
there. One might hope that, rather than leaving the universe, information
can be reflected off of the boundary line.
However there appears to be a severe problem with this: the effect
of throwing matter at the line $\phi=\phic$ is to change the
trajectory of the boundary from a timelike one to a spacelike one \RST.
No local boundary condition can alter this conclusion. It does not seem
possible to define sensible, unitary dynamics for a system with
spacelike boundaries.

\subsec { Sailing through the Singularity.}

Another possibility is that the boundary conditions give a prescription for
continuing through the singularity. It is well-known that there are
certain types of mild ``shock-wave'' singularities in general relativity
which do not prevent a unique and consistent evolution. The present
class of singularities are in fact quite mild -- neither the dilaton or
metric diverge, while their first derivatives typically blow up like
${t}^{-1/3}$ as the singularity is approached. There is a
substantial literature on the problem of continuing through singularities
arising in
PDE's of this general type.   It in fact appears
likely that it is possible to find a rule for continuing through
the singularity -- the difficulty is in keeping all the fields real while
doing so. Work on this issue is in progress.

\centerline{\bf ACKNOWLEDGMENTS}
We have been informed that related results
have been independently
obtained by S. Hawking and by L. Susskind and L. Thorlacius.
We are grateful to them for discussing their results with
us prior to publication, and to G. Horowitz for useful discussions.
This work
was supported in part by DOE grant DE-FG03-91ER40168, NSF grants
PHY90-00386, DMS91-04532 and by
NSF PYI grants, PHY-9157463 to S.B.G. and PHY-9196117 to J.A.H.
\centerline{\bf NOTE ADDED}
Hawking's work has appeared in ``Evaporation of Two Dimensional
Black Holes'' Caltech preprint CALT-68-1774 and hepth@xxx/9203052,
and Susskind and Thorlacius' work has appeared in
``Hawking Radiation and Back-Reaction'' Stanford preprint
SU-ITP-92-12 and hepth@xxx/9203054.

\listrefs

\vfil\eject

\centerline{\bf Figure Captions}
\bigskip
\item{~~~Fig.~1a.} A plot of $\phi$ versus $\sigma$ for quantum kinks of
mass $M=\lambda, 10\lambda,
50\lambda$ for $\phic=-2$. Integrating in from infinity, the solutions closely
resemble
the linear dilaton vacuum until $\phic$ is reached, at which point the
solutions bounce back towards weak coupling at minus infinity.

\item {Fig.~1b.} A plot of $\rho$ versus $\sigma$ for quantum kinks of
mass $M=\lambda, 10\lambda,
50\lambda$ for $\phic=-2$. Integrating in from infinity,
$\rho$ is nearly zero until the bounce occurs. It asymptotes to
minus infinity at $\sigma=-\infty$ with a constant linear slope plus
logarithmic corrections.
This implies that $\sigma=-\infty$ is a finite distance away, and that
there is a horizon there.

\item{Fig.~1c.} A plot of $\phi$ versus $s=e^{2\lambda\sigma}$ for quantum
kinks of
mass $M=\lambda, 10\lambda,
50\lambda$ for $\phic=-2$. The horizon is pulled into $s=0$ in these
coordinates, and it is evident that the dilaton goes to zero coupling
there.

\item{Fig.~1d.} A plot of $\rho$ versus $s=e^{2\lambda\sigma}$ for quantum
kinks of
mass $M=\lambda, 10\lambda,
50\lambda$ for $\phic=-2$. $\rho$ diverges on the horizon.

\item{Fig.~2a.} A plot of $\phi$ versus $s=e^{2\lambda\sigma}$ for a black hole
in equilibrium with Hawking radiation. The initial conditions are chosen so
as to ensure regularity at the horizon ($s=0$). Inside the horizon,
where $s<0$ is a timelike coordinate, $\phi$ rapidly increases and
reaches a singularity at $\phic=-2$ in finite time.

\item{Fig.~2b.} A plot of $\rho$ versus $s$ for a black hole in equilibrium
with Hawking radiation.

\end